# Reduce Noise in Computed Tomography Image using Adaptive Gaussian Filter

Rini Mayasari*, Nono Heryana**

*Fakultas Ilmu Komputer, Universitas Singaperbangsa Karawang, Indonesia

**Fakultas Ilmu Komputer, Universitas Singaperbangsa Karawang, Indonesia

----------------------------------------************************--------------------------------

## Abstract:

One image processing application that is very helpful for humans is to improve image quality, poor image quality makes the image more difficult to interpret because the information conveyed by the image is reduced. In the process of the acquisition of medical images, the resulting image has decreased quality (degraded) due to external factors and medical equipment used. For this reason, it is necessary to have an image processing process to improve the quality of medical images, so that later it is expected to help facilitate medical personnel in analyzing and translating medical images, which will lead to an improvement in the quality of diagnosis. In this study, an analysis will be carried out to improve the quality of medical images with noise reduction with the Gaussian Filter Method. Next, it is carried out, and tested against medical images, in this case, the lung photo image. The test image is given noise in the form of impulse salt & pepper and adaptive Gaussian then analyzed its performance qualitatively by comparing the output filter image, noise image, and the original image by naked eye.

----------------------------------------************************--------------------------------

## I. INTRODUCTION

Image processing is processed image, especially by using a computer, to be a better quality image [1]. In the process, many image processing involves visual perception and has the characteristics of input data and output information in the form of digital image files [2].

The medical image of the scanning results, in the form of a gray level digital image, has decreased quality (degraded) due to external factors and medical equipment used so that the utilization of image processing is perceived as not optimal [3]. Thus the process of improving the quality of medical images must also use gray level image processing techniques [4]. Medical images in the field of medicine require analysis with a high degree of accuracy, especially in diagnosing certain diseases such as cancer. The high mortality rate caused by cancer is considered in the selection of lung cancer as a test image, accounting for 8.8 million deaths in 2015 from the official WHO (World Health Organization) website, lung cancer is the most common cause of death cancer is 1.69 million deaths, Liver (788,000 deaths), Colorectal (774,000 deaths), Stomach (754,000 deaths), Breast (571,000 deaths).

The decline in the quality of image quality or degradation, for example, containing defects or noise, the color is too contrasting, less sharp, blurred, and so on makes the image more difficult to interpret because the information conveyed by the image is reduced [1]. The degradation in question includes noise (which is an error in pixel values) [5].

In some applications such as medical image processing and remote sensing through satellite imagery, noise reduction is an important requirement that must be done before the next processing stage. This phase is a preprocessing stage that must be done to improve image quality (image enhancement) [6].

Noise or noise is an image or pixel that interferes with image quality. Noise can be caused by physical (optical) interference on the acquisition device or intentionally due to inappropriate processing. Examples are black or white spots that appear randomly unwanted in images or white spots that are normally distributed. Noise with random spots is called noise (salt & pepper) and noise with normal white spots is called Gaussian noise [7]. Disruptive noise because it reduces image quality during printing, makes it difficult to identify criminals on CCTV images, also makes it difficult to detect cancer cells or diseases in medical images (MRI, CTScan, XRay) [6].

Responding to this, it is necessary to do an image repair process using the image filtering method. By using filtering techniques will be carried out the process of noise reduction in medical images which previously has been given a combined noise and salt and pepper. Image processing in this study consists of the process of image input, image conversion, noise addition, image reduction/noise reduction with the Gaussian filter method.

Based on the research that has been done by Dimas Ari [7] shows that the Gaussian Filter method has better effectiveness than the Wiener Filter in reducing images that contain a combination of Gaussian noise and salt and pepper.

From this background, this research will analyze the Gaussian filter method by reducing the combined two noise, which is by combining Gaussian noise and salt and pepper noise which will be tested with the Gaussian filter method.

## II. METHOD

The research method used in the study of noise reduction on medical images with the Gaussian filter method was carried out in accordance with the research flow as shown in figure 1 below:

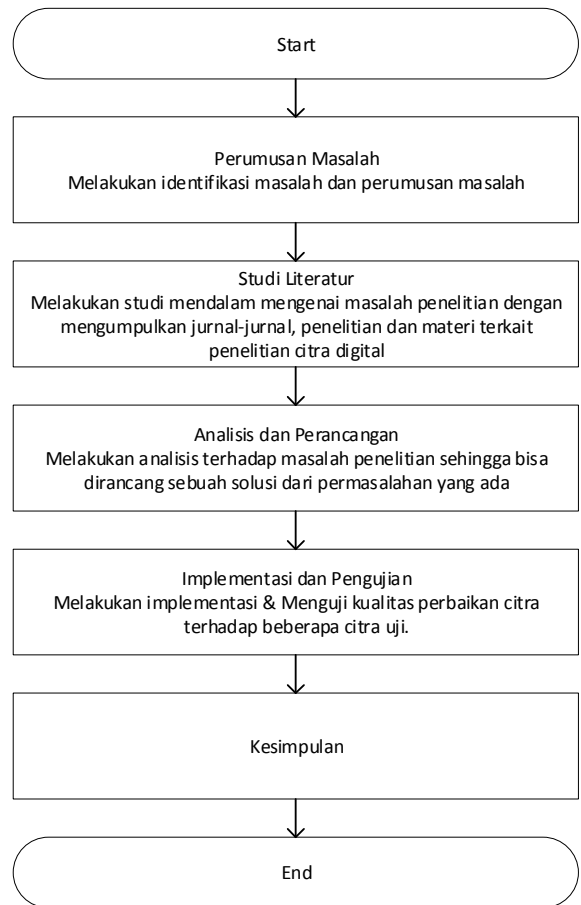

Figure 1 Research Stage

## III. RESULTS

Test data used in this research is .jpg each image is 512x512 pixels. The medical image that will be processed in this study is the image taken using Computed Tomography Scan (CT Scan) lung technology. This technique uses digital images of 30 images of lung cancer. Trial image data in the form of chest x-ray images taken from a common dataset of JSRT (Japanese Society of Radiological Technology). The trial was conducted to find out whether the program can run according to the implementation scenario of the Gaussian filter method.

The implementation of the Gaussian filter method can be used in the field of image analysis, especially for the process of avoiding noise. In a Gaussian filter, the intensity value of each pixel is replaced by the average of the weighting values for each neighbor's pixels and the pixel itself. Neighboring pixels are pixels around the pixel in question. The number of neighbours involved depends on the filter designed.

An input image is as many as 10 images with the addition of noise salt and paper, then noise removal is done on each input image using the Gaussian filter method.

Based on the results of the scenario implementation carried out in this study, according to the flow belowin figure 2.

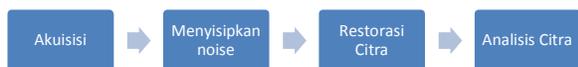

a. Data Acquisition

Data acquisition systems can be defined as a system that functions to retrieve, collect and prepare data, to process it to produce the desired data. In image-type data that is still in the form of an analog, the image must be represented numerically with discrete values so that it can be processed with a digital computer.

In this study the image used is the CT-scan image of lung cancer as shown below:

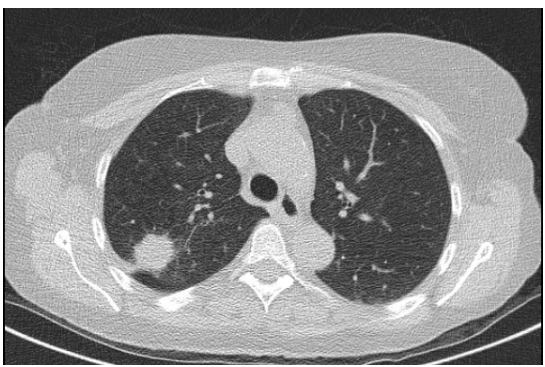

Figure 2 Original Image

b. Added Noise

Noise is an unwanted signal that enters and blends with the signal to be processed. The noise comes from various sources, both from the circuit and from the outside environment.

Noise in the image does not only occur due to imperfections in the process of taking pictures or during the transmission process. But also because of the impurities that occur in an image. There is some noise that can be attached to an image and one of them is Salt and Pepper noise in the form of black or white dots scattered in an image.

In this study, the scenario of the amount of noise from the image that will be restored is about 10% of the total area of the image.

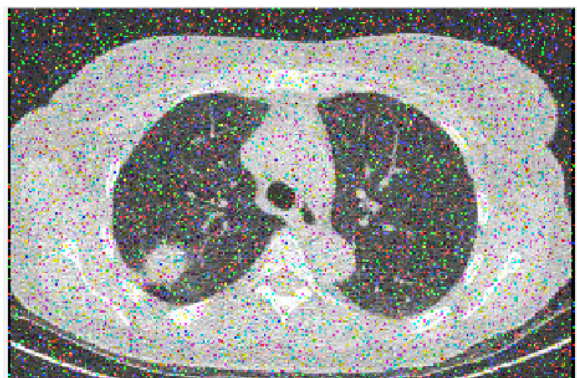

Figure 3 Noise Image with Salt & Pepper Noise Ratio 10%

In the image restoration stage using image filtering technique. Image Filtering is a technique used to modify the image so as to make the image look better in terms of image quality so that it is easier to analyze.

In this study, an adaptive Gaussian filter method is used to reduce the amount of noise in the ct scan of lung cancer. However, the results of image restoration using an adaptive Gaussian filter is as shown below in figure 4.

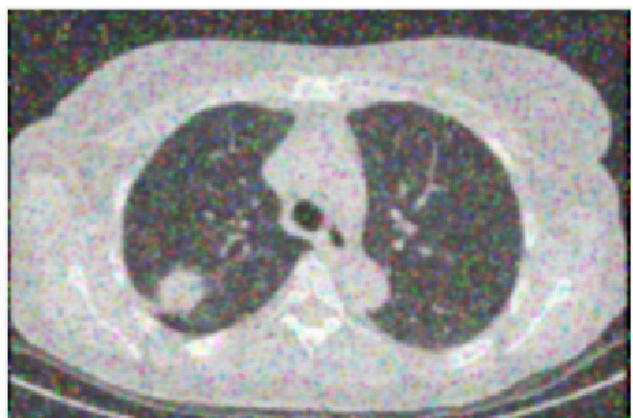

Figure 4 Restoration Image

c. Image Analysis

Peak Signal to Noise Ratio (PSNR) is the ratio between the maximum value of a signal measured by the amount of noise that affects the signal.

PSNR is usually measured in decibels (dB). PSNR is used to compare the quality of the cover image before and after the message is inserted. To determine the PSNR, the MSE value (Mean Square Error) must first be determined.

MSE is the average square error value between the original image and the manipulation image, MSE is the average square error value between the original image and the restoration image.

In addition to determining MSE, SNR, and PSNR, in this study also used the Structural Similarity Index (SSIM) to measure the quality of the analyzed images.

As for the results of the comparison of the analysis of the noise image and the results of the restoration is in the table below:

**Table 1 Analisis Citra Bernoise**

| No | Nama file | Perbandingan Nilai | | | |
|----|-----------|------|------|------|------|
|    |           | MSE | SNR | PSNR | SSIM |
| 1  | Uji1.jpg  | 2085.7967 | 10.5073 | 14.9381 | 0.3224 |
| 2  | Uji2.jpg  | 2417.0623 | 9.5459  | 14.2979 | 0.1705 |
| 3  | Uji3.jpg  | 2397.1857 | 10.9197 | 14.3338 | 0.2607 |
| 4  | Uji4.jpg  | 2106.5978 | 10.0909 | 14.8950 | 0.1388 |
| 5  | Uji5.jpg  | 2119.1315 | 10.0720 | 14.8692 | 0.1383 |
| 6  | Uji6.jpg  | 2085.1872 | 10.1001 | 14.9381 | 0.3256 |
| 7  | Uji7.jpg  | 2132.9764 | 10.0923 | 14.8803 | 0.1346 |
| 8  | Uji8.jpg  | 2267.5321 | 10.8175 | 14.3467 | 0.2564 |
| 9  | Uji9.jpg  | 2598.4572 | 9.7862  | 14.3562 | 0.2601 |
| 10 | Uji10.jpg | 2365.2462 | 10.8451 | 14.6528 | 0.1435 |

**Table 2 Analisis Citra Hasil Restorasi**

| No | Nama file | Perbandingan Nilai | | | |
|----|-----------|------|------|------|------|
|    |           | MSE | SNR | PSNR | SSIM |
| 1  | Uji1.jpg  | 1761.8828 | 10.7933 | 15.6710 | 0.3788 |
| 2  | Uji2.jpg  | 1828.0268 | 10.2613 | 15.5110 | 0.3777 |
| 3  | Uji3.jpg  | 2010.0982 | 11.2728 | 15.0986 | 0.3966 |
| 4  | Uji4.jpg  | 1590.5220 | 10.8848 | 16.1154 | 0.3601 |
| 5  | Uji5.jpg  | 1600.9984 | 10.8614 | 16.0869 | 0.3593 |
| 6  | Uji6.jpg  | 1754.6792 | 10.7562 | 15.6427 | 0.3468 |
| 7  | Uji7.jpg  | 2142.5638 | 11.3257 | 15.0933 | 0.3968 |
| 8  | Uji8.jpg  | 2009.7831 | 10.8694 | 16.0832 | 0.3732 |
| 9  | Uji9.jpg  | 1902.8752 | 10.0744 | 15.2242 | 0.3782 |
| 10 | Uji10.jpg | 2008.8392 | 10.2574 | 15.5142 | 0.3775 |

Based on the results of the comparison above, it can be seen that the results of the restoration image analysis have MSE, SNR, PSNR and SSIM values having values close to the calculation value in the original image.

### IV. CONCLUSIONS

Based on the results of testing the ct-scan image of lung cancer, it can be concluded that the PSNR value is influenced by the MSE value of the type tested. And from the results of quality measurements using the Structural Similarity Index (SSIM) the higher the SSIM value the better the quality.

### ACKNOWLEDGMENT

This research was funded by an internal research grant (DIPA UNSIKA) from Universitas Singaperbangsa Karawang.